\RequirePackage{fix-cm}
\documentclass[smallextended]{svjour3}       
\usepackage{graphicx}
\usepackage{amsmath}

\begin{document}

\title{An algorithm for discrete fractional Hadamard transform 
}

\titlerunning{Novel DFRHT algorithm}        

\author{Aleksandr Cariow \and 
        Dorota Majorkowska-Mech}


\institute{A. Cariow \at
              Faculty of Computer Science and Information Technology, West Pomeranian University of Technology, Szczecin, Poland \\
              Tel.: +48-91-4495573\\
              Fax: +48-91-4495559\\
              \email{atariov@wi.zut.edu.pl}           
           \and
            D. Majorkowska-Mech \at
              Faculty of Computer Science and Information Technology, West Pomeranian University of Technology, Szczecin, Poland\\
              Tel.: +48-91-4495582\\
              Fax: +48-91-4495559\\
              \email{dmajorkowska@wi.zut.edu.pl}   
}

\date{Received: date / Accepted: date}

\maketitle

\begin{abstract}
We present a novel algorithm for calculating the discrete fractional Hadamard transform for data vectors whose size $N$ is a power of two. A direct method for calculation of the discrete fractional Hadamard transform requires $N^2$ multiplications, while in proposed algorithm the number of real multiplications is reduced to $N$log$_2N$.

\end{abstract}

\section{Introduction}
\label{intro}
Discrete fractional transforms are the generalizations of the ordinary discrete transforms with one additional fractional parameter. In the past decades, various discrete fractional transforms including discrete Fourier transform \cite{pe0}, \cite{can}, discrete fractional Hartley transform \cite {pe2}, discrete fractional cosine transforms and discrete sine transform \cite {pe1} have been introduced and found wide applications in many scientific and technological areas including digital signal processing \cite {oza}, image encryption \cite {hen}, \cite{liu}, \cite{nis} and digital watermarking \cite{dju} and others. Different fast algorithms for their implementations have been separately developed to minimize computational complexity and implementation costs. In  \cite{pe3} a discrete fractional Hadamard transform for the vector of length $N\! =\! 2^n$ was introduced, however a fast algorithm for the realization of this transform has not been proposed. In our previous paper \cite{ma} we describe a rationalized algorithm for DFRHT possessing a reduced number of multiplications and additions. Analysis of the mentioned algorithm shows that not all of existing improvement possibilities have been realized. In this paper, we proposed a novel algorithm for the discrete fractional Hadamard transform that require fewer total real additions and multiplications than our previously published solution.

\section{Mathematical background}
\label{sec:1}
A Hadamard matrix of order $N$ is a $N\times N$ symmetric matrix whose entries are either 1 or $-1$ and whose rows are mutually orthogonal. In this paper we will use the normalized form of this matrix and we will denote it by $\mathbf{H}_N$. For $N=2^n$ the Hadamard matrices can be recursively obtained due to Sylvester's construction \cite{syl}:
\begin{equation}
\mathbf{H}_2=\frac{1}{\sqrt{2}}
\left[
\begin{array}{cc}
1 & 1  \\
1 & -1 \\
\end{array} \right]\!,\
\mathbf{H}_{N}=\frac{1}{\sqrt{2}}
\left[
\begin{array}{cc}
\mathbf{H}_{\frac{N}{2}} & \mathbf{H}_{\frac{N}{2}}\\
\mathbf{H}_{\frac{N}{2}}& -\mathbf{H}_{\frac{N}{2}}
\end{array}\right]
\end{equation}
for $N=4,8,\ldots,2^n$.

Definition of the discrete fractional Hadamard (DFRHT) transform is based on an eigenvalue decomposition of the DHT matrix. Any real symmetric matrix (including the Hadamard matrix) can be diagonalized, e.g. written as a product \cite{kor}

\begin{equation}
\mathbf{H}_N=\mathbf{Z}_N \mathbf{\Lambda}_N \mathbf{Z}_N^T=\sum\limits_{k=0}^{N-1}\lambda_k \mathbf{z}_N^{(k)} (\mathbf{z}_N^{(k)})^T \label{eq:rozklad}
\end{equation}
where $\mathbf{\Lambda}_N$ is a diagonal matrix of order $2^n$, whose diagonal entries are the eigenvalues of $\mathbf{H}_N$

\begin{equation}
\mathbf{\Lambda}_N=
\left[\begin{array}{cccc}
\lambda_0 & & & \\
 &\lambda_1& & \mathbf{0}\;\;\;\\
 \;\;\;\mathbf{0}&& \ddots& \\
 & & &\lambda_{N-1}\\
\end{array} \right]
\end{equation}\\
$\mathbf{Z}_N=[\mathbf{z}_{N}^{(0)}\mid\mathbf{z}_{N}^{(1)}\mid\ldots\mid\mathbf{z}_{N}^{(N-1)}]$ - the matrix whose columns are normalized mutually orthogonal eigenvectors of the matrix $\mathbf{H}_N$. The eigenvector $\mathbf{z}_{N}^{(k)}$ is related to the eigenvalue $\lambda_{k}$. A superscript $T$ denotes the matrix transposition. 

The DFRHT matrix of order $N=2^n$ with real parameter $\alpha$ was first defined in  \cite{pe3}. This matrix can be regarded as a power of the DHT matrix, where the exponent $a\!=\!\alpha/\pi$
\begin{equation}
\mathbf{H}_{N}^{a}=\mathbf{Z}_N \mathbf{\Lambda}_{N}^{a}\mathbf{Z}_N^T=\sum\limits_{k=0}^{N-1}\lambda_k^a \mathbf{z}_N^{(k)} (\mathbf{z}_N^{(k)})^T \label{eq:rozklad2}
\end{equation} 
 For $a\! =\! 0$ the DFRHT matrix is converted into the identity matrix, and for $a \!\!=\! \!1$ it is transformed into the ordinary DHT matrix. Generally the DFRHT matrix is complex-valued.

An essential operation, by obtaining the discrete fractional Hadamard matrix, defined by (\ref{eq:rozklad2}), is calculating the eigenvalues and the eigenvectors of the matrix $\mathbf{H}_N$. The only eigenvalues of the unnormalized Hadamard matrix of order $N\!=\!2^n$ are known to be 
$2^{n/2}$ and $-2^{n/2}$ \cite{yar1}, hence the normalized Hadamard matrix $\mathbf{H}_{N}$ has only the eigenvalues 1 and $-1$. A method for finding the eigenvectors of Hadamard matrix was firstly presented in \cite{yar}, but in \cite{pe3} a recursive method for calculation the eigenvectors of the Hadamard matrix order $2^{n+1}$ based on the eigenvectors of the Hadamard matrix of order $2^n$ has been proposed. We will use this method to obtain the DFRHT matrix. Here we will present it briefly. 

In \cite{yar} it was proven that if $\mathbf{v}_{N}^{(k)}$ ($k\! =\! 0,1,\ldots,N-1$) is an eigenvector of Hadamard matrix of order $N\!=\!2^n$ associated with an eigenvalue $\lambda$, then vector 
\begin{equation}
\mathbf{\hat v}_{2N}^{(k)}=
\left[
\begin{array}{c}
\mathbf{v}_N^{(k)}\\
(\sqrt{2}-1) \mathbf{v}_{N}^{(k)}\\
\end{array} \right] \label{eq:wektorhat}
\end{equation} 
will be an eigenvector of the matrix $\mathbf{H}_{2N}$ associated with the eigenvalue $\lambda$. \newline
In \cite{pe3} it was proven that if $\mathbf{v}_{N}^{(k)}$ is an eigenvector of $\mathbf{H}_N$ associated with an eigenvalue $\lambda$, then the vector
\begin{equation}
\mathbf{\tilde{v}}_{2N}^{(k)}=\left[
\begin{array}{c}
(1-\sqrt{2}) \mathbf{v}_{N}^{(k)}\\
\mathbf{v}_{N}^{(k)}\\
\end{array} \right] \label{eq:wektortilde}
\end{equation} 
will be an eigenvector of the matrix $\mathbf{H}_{2N}$ associated with the eigenvalue $-\lambda$. \newline
These two results allow as to generate the eigenvectors of Hadamard matrix of order $2^{n+1}$  from the eigenvectors of Hadamard Matrix of order $2^n$. Knowing the straightforward calculated eigenvectors of the matrix $\mathbf{H}_2$ 
\begin{equation}
\mathbf{v}_{2}^{(0)}=
\left[
\begin{array}{c}
1\\
\sqrt{2}-1\\
\end{array} \right]\;\;
\mathbf{v}_{2}^{(1)}=\left[
\begin{array}{c}
1-\sqrt{2}\\
1\\
\end{array} \right] \label{eq:wektoryh2}
\end{equation}
associated with eigenvalues 1 and $-1$ respectively, the eigenvectors for Hadamard matrix of arbitrary order $N\!=\!2^n$ can be recursively computed.  In \cite{pe3} it was also shown so this recursively computed eigenvectors of matrix $\mathbf{H}_{N}$ will be orthogonal. It should be noted that for any $N\!=\!2^n$ there are only two distinct eigenvalues of Hadamard matrix, so for $N\geq 4$ the eigenvalues are degenerated.  Because of this fact the set of eigenvectors proposed in \cite{yar} and \cite {pe3} is not unique.
The igenvectors $\mathbf{v}_{N}^{(k)}$ for $k=0,1,\ldots,N-1$, which are columns of the matrix $\mathbf{Z}_N$ (after normalization), as well as their associated eigenvalues $\lambda_k$, can be however ordered in different ways. In \cite {pe3} it has been also established a method of ordering the eigenvectors. In many cases, including the case of discrete fractional transforms is used so-called sequency ordering of the eigenvectors. This means that the $k$-th eigenvector has $k$ sign-changes. The discrete Hermite-Gaussians, eigenvectors of discrete Fourier transform matrix are ordered this way as well \cite{can}. We will show this method of ordering of the eigenvectors in Example \ref{Example1}.

\begin{example}\label{Example1}
The number of sign-changes  in  eigenvectors $\mathbf{v}_{2}^{(0)}$ and $\mathbf{v}_{2}^{(1)}$ of matrix $\mathbf{H}_2$, determined by (\ref{eq:wektoryh2}), is equal to 0 and 1 respectively. Using expressions (\ref{eq:wektorhat}) and (\ref{eq:wektortilde}) we obtain the eigenvectors of matrix $\mathbf{H}_4$:

\begin{displaymath}
\mathbf{\hat v}_{4}^{(0)}=
\left[
\begin{array}{c}
1\\
b \\
b\\
b^2\\
\end{array} \right]\;\;
\mathbf{\tilde v}_{4}^{(0)}=
\left[
\begin{array}{c}
-b\\
-b^2 \\
1\\
b\\
\end{array} \right]\;\;
\mathbf{\hat v}_{4}^{(1)}=
\left[
\begin{array}{c}
-b\\
1 \\
-b^2\\
b\\
\end{array} \right]\;\;
\mathbf{\tilde v}_{4}^{(1)}=
\left[
\begin{array}{c}
b^2\\
-b \\
-b\\
1\\
\end{array} \right],
\end{displaymath}
where $b\! =\! \sqrt{2}-1$. The numbers of sign-changes in the above vectors are 0, 1, 3, 2 respectively ($b>0$). Therefore, a sequency ordered set of eigenvectors of matrix $\mathbf{H}_4$ will be as follows:

\begin{displaymath}
\mathbf{v}_{4}^{(0)}=\mathbf{\hat v}_{4}^{(0)}\;\;
\mathbf{v}_{4}^{(1)}=\mathbf{\tilde v}_{4}^{(0)}\;\;
\mathbf{v}_{4}^{(2)}=\mathbf{\tilde v}_{4}^{(1)}\;\;
\mathbf{v}_{4}^{(3)}=\mathbf{\hat v}_{4}^{(1)}.
\end{displaymath}
The corresponding eigenvalues will be equal to:
\begin{displaymath}
\lambda_0=1\;\;
\lambda_1=-1\;\;
\lambda_2=1\;\;
\lambda_3=-1.
\end{displaymath}

\end{example}
The relations obtained in Example \ref{Example1} can be easily generalized as follows:

\begin{equation}
\left\{ \begin{array}{l}
\mathbf{v}_{2N}^{(4l)}=\mathbf{\hat v}_{2N}^{(2l)}\\[6pt]
\mathbf{v}_{2N}^{(4l+1)}=\mathbf{\tilde v}_{2N}^{(2l)}\\[6pt]
\mathbf{v}_{2N}^{(4l+2)}=\mathbf{\tilde v}_{2N}^{(2l+1)}\\[6pt]
\mathbf{v}_{2N}^{(4l+3)}=\mathbf{\hat v}_{2N}^{(2l+1)}
\end{array} \right.
\end{equation}
for $l=0,1,\ldots,\frac{N}{2}-1$ and
\begin{equation}
\lambda_k=(-1)^k.
\end{equation}
for $k=0,1,\ldots,2N-1$.
 
Both the eigenvectors of the matrix $\mathbf{H}_2$ and  the eigenvectors obtained for higher order Hadamard matrices are not normalized. Let the notation $\big \Vert \mathbf{v}_{N}^{(k)} \big \Vert$ means the Euclidean norm of vector $\mathbf{v}_{N}^{(k)}$. In \cite {ma} it was shown that for any $N=2^n$ we have the relationship
\begin{equation}
\big \Vert \mathbf{v}_{N}^{(k)} \big \Vert^2=\big (1+b^2 \big)^n
\end{equation}
where $k=0,1,\ldots,N-1$ and $b=\sqrt{2}-1$. If we take the designation $c\! =\! 1+b^2$, then normalized eigenvectors of Hadamard matrix of order $N=2^n$ will take the form
\begin{equation}
\mathbf{z}_{N}^{(k)}=\frac{\mathbf{v}_{N}^{(k)} }{\big\Vert \mathbf{v}_{N}^{(k)}\big\Vert }=\frac{\mathbf{v}_{N}^{(k)} }{\sqrt{c^n}}
\end{equation}
Using the normalized and sequency ordered eigenvectors of the Hadamard matrix, the eigenvalue decomposition (\ref{eq:rozklad}) of the Hadamard matrix can be written as follows: 

\begin{equation}
\mathbf{H}_N=\mathbf{Z}_N \mathbf{\Lambda}_N \mathbf{Z}_N^T=\frac{1}{c^n}\mathbf{V}_N
\mathbf{\Lambda}_N \mathbf{V}_N^T  
\end{equation}
where $\mathbf{\Lambda}_N$ is the diagonal matrix whose non-zero elements are
\begin{equation}
\lambda_k=(-1)^k=e^{-jk\pi}
\end{equation}
for $k=0,1,\ldots,N-1$. Hence the definition (\ref{eq:rozklad2}) of DFRHT matrix will take the form:
\begin{equation}
\mathbf{H}_{N}^{a}=
\frac{1}{c^n}\mathbf{V}_N \mathbf{\Lambda}_{N}^{a}\mathbf{V}_N^T \label{eq:rozklad2n}
\end{equation} 
where
\begin{equation}
\lambda_{k}^{a}=e^{-jk \pi a}
\end{equation}
for $k=0,1,\ldots,N-1$.

Our goal is to calculate the discrete fractional Hadamard transform for an input signal $\mathbf{x}_{N}$ in which the number of samples is equal to $N\!=\!2^n$. By $\mathbf{y}_{N}^{(a)}$ we denote an output signal which is calculated using the formula
\begin{equation}
\mathbf{y}_{N}^{(a)}=\mathbf{H}_{N}^{a}\mathbf{x}_{N} \label{eq:transformata}
\end{equation}
Supposing that the matrix $\mathbf{H}_{N}^{a}$ is given, to calculate the output signal it is necessary to perform $N^2$ complex multiplications and $N(N-1)$ complex additions. If the input signal is real, then the number of real multiplications will be equal to $2N^2$, and the number of real additions will be equal to $2N(N-1)$.

If we use the decomposition (\ref{eq:rozklad2n}) of the matrix $\mathbf{H}_{N}^{a}$ by calculating (\ref{eq:transformata}) and will perform the matrix-vector multiplication from the right side to the left, the most time-consuming operations are multiplications of matrices $\mathbf{V}_N^T$ and $\mathbf{V}_N$ by the vector, because those matrices are not diagonal. If we interchange the columns of the matrix  $\mathbf{V}_N$ in the prescribed manner, we obtain a matrix $\mathbf{\overline{V}}_N$ of special structure, which can be generated recursively. We will show it in Example \ref{Example2}. 
It will allow to reduce the number of arithmetical operations by calculating the products of matrices $\mathbf{V}_N^T$ and $\mathbf{V}_N$ by the vector.

\begin{example}\label{Example2}
The matrices $\mathbf{V}_N$  for $N=2,4,8$ are as follows:
\begin{displaymath}
\mathbf{V}_2=
\left[
\begin{array}{cc}
1&-b\\
b&1\\
\end{array} \right],\;\; 
\mathbf{V}_4=
\left[
\begin{array}{cccc}
1&-b&b^2&-b\\
b&-b^2&-b&1\\
b&1&-b&-b^2\\
b^2&b&1&b\\
\end{array} \right],\;\;
\end{displaymath}
\begin{displaymath}
\mathbf{V}_8=
\left[
\begin{array}{cccccccc}
1&-b&b^2&-b&b^2&-b^3&b^2&-b\\
b&-b^2&b^3&-b^2&-b&b^2&-b&1\\
b&-b^2&-b&1&-b&b^2&b^3&-b^2\\
b^2&-b^3&-b^2&b&1&-b&-b^2&b\\
b&1&-b&-b^2&b^3&b^2&-b&-b^2\\
b^2&b&-b^2&-b^3&-b^2&-b&1&b\\
b^2&b&1&b&-b^2&-b&-b^2&-b^3\\
b^3&b^2&b&b^2&b&1&b&b^2\\
\end{array} \right].
\end{displaymath}
The matrix  $\mathbf{V}_2$ has some specific structure. Now we consider the matrix  $\mathbf{V}_4$. If in the matrix  $\mathbf{V}_4$ the second and fourth columns will be interchange and then the third and fourth columns will be interchange too, we obtain the following matrix:
\begin{displaymath}
\mathbf{\overline{V}}_4=
\left[
\begin{array}{cccc}
1&-b&-b&b^2\\
b&1&-b^2&-b\\
b&-b^2&1&-b\\
b^2&b&b&1\\
\end{array} \right]
=
\left[
\begin{array}{cc}
\mathbf{V}_2&-b\mathbf{V}_2\\ 
b\mathbf{V}_2&\mathbf{V}_2\\
\end{array} \right].
\end{displaymath}
The matrix $\mathbf{V}_4$ differs from the matrix $\mathbf{\overline{V}}_4$ only in the order of the columns. Therefore, the matrix $\mathbf{V}_4$ can be obtained by post-multiplying the matrix $\mathbf{\overline{V}}_4$ by the permutation matrix $\mathbf{P}_4$:
\begin{displaymath}
\mathbf{V}_4=\mathbf{\overline{V}}_4\mathbf{P}_4,
\end{displaymath}
where
\begin{displaymath}
\mathbf{P}_4=
\left[
\begin{array}{cccc}
1&0&0&0\\
0&0&0&1\\
0&1&0&0\\
0&0&1&0\\
\end{array} \right].
\end{displaymath}

Now we consider the matrix $\mathbf{V}_8$. If we perform the following permutation of columns of this matrix:
\begin{displaymath}
\left(
\begin{array}{cccccccc}
1\hspace{0.2cm}&2\hspace{0.2cm}&3\hspace{0.2cm}&4\hspace{0.2cm}&5\hspace{0.2cm}&6\hspace{0.2cm}&7\hspace{0.2cm}&8\\
1\hspace{0.2cm}&8\hspace{0.2cm}&4\hspace{0.2cm}&5\hspace{0.2cm}&2\hspace{0.2cm}&7\hspace{0.2cm}&3\hspace{0.2cm}&6\\
\end{array} \right),
\end{displaymath}
as a result we obtain the following matrix:

\begin{displaymath}
\mathbf{\overline{V}}_8=
\left[
\begin{array}{cccccccc}
1&-b&-b&b^2&-b&b^2&b^2&-b^3\\
b&1&-b^2&-b&-b^2&-b&b^3&b^2\\
b&-b^2&1&-b&-b^2&b^3&-b&b^2\\
b^2&b&b&1&-b^3&-b^2&-b^2&-b\\
b&-b^2&-b^2&b^3&1&-b&-b&b^2\\
b^2&b&-b^3&-b^2&b&1&-b^2&-b\\
b^2&-b^3&b&-b^2&b&-b^2&1&-b\\
b^3&b^2&b^2&b&b^2&b&b&1\\
\end{array} \right]=
\left[
\begin{array}{cc}
\mathbf{\overline{V}}_4&-b\mathbf{\overline{V}}_4\\ 
b\mathbf{\overline{V}}_4&\mathbf{\overline{V}}_4\\
\end{array} \right].
\end{displaymath}
As previously, we can write:
\begin{displaymath}
\mathbf{V}_8=\mathbf{\overline{V}}_8\mathbf{P}_8,
\end{displaymath}
where
\begin{displaymath}
\mathbf{P}_8=
\left[
\begin{array}{cccccccc}
1&0&0&0&0&0&0&0\\
0&0&0&0&0&0&0&1\\
0&0&0&1&0&0&0&0\\
0&0&0&0&1&0&0&0\\
0&1&0&0&0&0&0&0\\
0&0&0&0&0&0&1&0\\
0&0&1&0&0&0&0&0\\
0&0&0&0&0&1&0&0\\
\end{array} \right].
\end{displaymath}
\end{example}
If we generalize the above considerations for $N=2^n$ we can write:
\begin{equation}
\mathbf{V}_N=\mathbf{\overline{V}}_N\mathbf{P}_N
\label{eq:permutacja}
\end{equation}
for $N=2,4,\ldots,2^n$. For $N=2$ we can also write
\begin{displaymath}
\mathbf{V}_2=\mathbf{\overline{V}}_2\mathbf{P}_2=\mathbf{\overline{V}}_2,
\end{displaymath}
where $\mathbf{P}_2$ is an identity matrix of order two
\begin{displaymath}
\mathbf{P}_2=\mathbf{I}_2.
\end{displaymath}
The permutation matrix $\mathbf{P}_{N}$ of order $N=2^{n}$ can be obtained recursively from the  permutation matrix $\mathbf{P}_{N/2}$ of order $2^{n-1}$  according to the following relation:

\begin{equation}
\mathbf{P}_2=
\left[
\begin{array}{cc}
1&0\\
0&1\\
\end{array} \right],\
\mathbf{P}_{N}=\mathbf{S}_{N}
\left[\begin{array}{cc}
\mathbf{P}_{\frac{N}{2}}\hspace{0.2cm} & \mathbf{0}_{\frac{N}{2}}\\
\mathbf{0}_{\frac{N}{2}}\hspace{0.2cm}& \mathbf{P}_{\frac{N}{2}}\mathbf{J}_{\frac{N}{2}}\\
\end{array}\right]\!
\end{equation}
for $N=4,8,\ldots,2^n$. $\mathbf{S}_{N}$ is the perfect shuffle permutation matrix of order $2^n$, $\mathbf{J}_{N/2}$ is the counter-identity matrix  of order $N/2$ and $\mathbf{0}_{N/2}$ is zero matrix. 
The perfect shuffle permutation is the permutation that splits the set consisting of an even number of elements into two piles and interleaves them. It can be written as follows:
\begin{displaymath}
\left(
\begin{array}{cccccc}
1\hspace{0.2cm}&2\hspace{0.2cm}&3\hspace{0.2cm}&4\hspace{0.2cm}&\ldots&\hspace{0.2cm}2n\\
1\hspace{0.2cm}&n\hspace{0.2cm}&2\hspace{0.2cm}&n+1\hspace{0.2cm}&\ldots&\hspace{0.2cm}2n\\
\end{array} \right).
\end{displaymath}
\newline
 For example

\begin{displaymath}
\mathbf{S}_8=
\left[
\begin{array}{cccccccc}
1&0&0&0&0&0&0&0\\
0&0&0&0&1&0&0&0\\
0&1&0&0&0&0&0&0\\
0&0&0&0&0&1&0&0\\
0&0&1&0&0&0&0&0\\
0&0&0&0&0&0&1&0\\
0&0&0&1&0&0&0&0\\
0&0&0&0&0&0&0&1\\
\end{array} \right], \
\mathbf{J}_4=
\left[
\begin{array}{cccc}
0&0&0&1\\
0&0&1&0\\
0&1&0&0\\
1&0&0&0\\
\end{array} \right].
\end{displaymath}

If we write the matrix $\mathbf{V}_N$ as a product  $\mathbf{\overline{V}}_N\mathbf{P}_{N}$  the expression (\ref{eq:rozklad2n}) will take the form: 

\begin{equation}
\mathbf{H}_{N}^{a}=\frac{1}{c^{n}}
\mathbf{\overline{V}}_N \mathbf{P}_N\mathbf{\Lambda}_{N}^{a} \mathbf{P}_N^T\mathbf{\overline{V}}_N^T \label{eq:rozklad3}
\end{equation} 
The product $\mathbf{P}_N\mathbf{\Lambda}_{N}^{a} \mathbf{P}_N^T$ is a diagonal matrix, which has the same diagonal entries as the matrix $\mathbf{\Lambda}_{N}^{a}$ but in different order and for a chosen parameter $a$ it may be prepared in advance. If we denote this product multiplied by a factor $1/c^n$ by $\mathbf{\tilde{\Lambda}}_{N}^{a}$:

\begin{equation}
\mathbf{\tilde{\Lambda}}_{N}^{a}=\frac{1}{c^{n}}\mathbf{P}_N\mathbf{\Lambda}_{N}^{a} \mathbf{P}_N^T.
\end{equation}
the DFRHT algorithm  (\ref{eq:transformata}) will take the following form:

\begin{equation}
\mathbf{y}_{N}^{(a)}=
\mathbf{\overline{V}}_N \mathbf{\tilde{\Lambda}}_{N}^{a} \mathbf{\overline{V}}_N^T \mathbf{x}_{N}  \label{eq:transformatan}
\end{equation}
where the matrix $\mathbf{\overline{V}}_N $ can be generated recursively: 

\begin{equation}
\mathbf{\overline V}_2=
\left[
\begin{array}{cc}
1&-b\\
b&1\\
\end{array} \right]\;\; 
\mathbf{\overline{V}}_{2k}=
\left[
\begin{array}{cc}
\mathbf{\overline{V}}_k&-b\mathbf{\overline{V}}_k\\ 
b\mathbf{\overline{V}}_k&\mathbf{\overline{V}}_k\\
\end{array} \right] \label{eq:defVN}
\end{equation}
for $k=2,4,\ldots,2^{n-1}$.

\section{Taking advantages of the particular structure of the matrix $\mathbf{\overline{V}}_N$}
\label{advantages}
\begin{figure}[h!]
\centering
\includegraphics[width=0.7\textwidth]{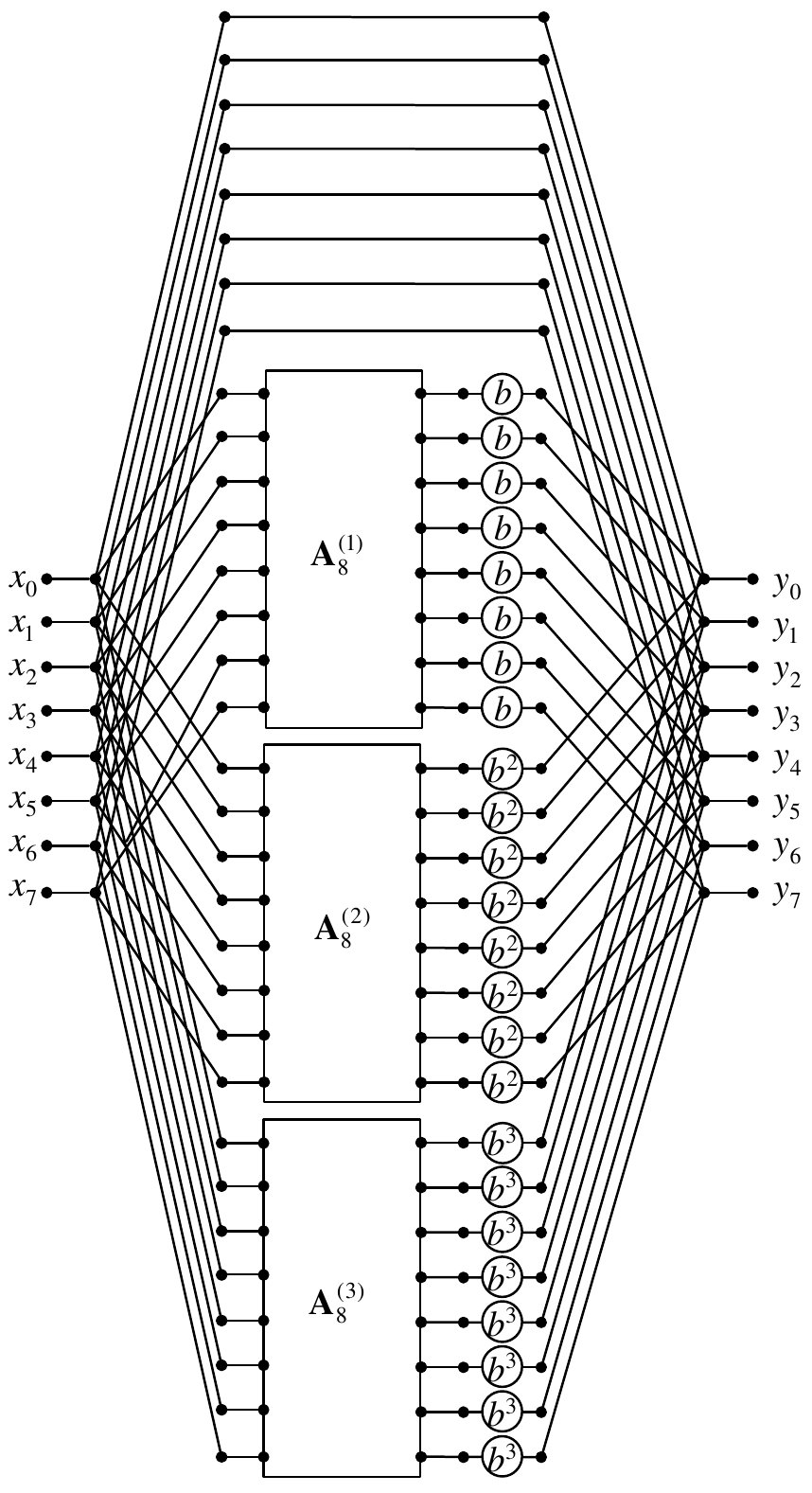} 
\caption{The graph-structural model of calculating the product $\mathbf{y}_8=\mathbf{\overline{V}}_8\mathbf{x}_8$}
\label{fig:figure1}  
\end{figure}

The most time-consuming operations by calculating the DFRHT transform according to (\ref{eq:transformatan})  are multiplications of matrices $\mathbf{\overline{V}}_N^T$ and $\mathbf{\overline{V}}_N$ by the vector. Since in the matrix $\mathbf{\overline{V}}_N$ occur only following powers of $b: b^n,b^{n-1},\ldots,b^0=1$ we can write this matrix as follows:
\begin{equation}
\mathbf{\overline V}_N=\mathbf{A}_N^{(0)}+b\mathbf{A}_N^{(1)}+b^2\mathbf{A}_N^{(2)}+\ldots+b^n\mathbf{A}_N^{(n)} \label{eq:sum}
\end{equation}
In the Figure \ref{fig:figure1} it was shown the way of calculating the matrix-vector product $\mathbf{y}_8=\mathbf{\overline{V}}_8\mathbf{x}_8$, using the expression (\ref{eq:sum}). 
In this paper, the graph-structural models and data flow diagrams are oriented from left to right. Straight lines in the figures denote the operation of data transfer. We use the usual lines without arrows specifically so as not to clutter the picture. Note that the circles in this figure shows the operations of multiplication by a number inscribed inside a circle.  In turn, the rectangles indicate the matrix-vector multiplications by matrices

Although it may seem strange, we will see that such an operation allows to reduce the number of multiplication and additions by multiplying the matrix $\mathbf{\overline V}_N$ by a vector.
It should be noted that because of the recursive relation (\ref{eq:defVN}) between matrices $\mathbf{\overline{V}}_{N}$ and $\mathbf{\overline{V}}_{N/2}$, the following recursive relation between the matrices $\mathbf{A}_{N}^{(k)}$,  $\mathbf{A}_{N/2}^{(k)}$ and  $\mathbf{A}_{N/2}^{(k-1)}$ occurs:
\begin{displaymath}
\mathbf{A}_{N}^{(0)}=
\left[
\begin{array}{cc}
\mathbf{A}_{N/2}^{(0)}&\mathbf{0}_{N/2}\\
\mathbf{0}_{N/2}&\mathbf{A}_{N/2}^{(0)}\\
\end{array} \right]=\mathbf{I}_{N}
\end{displaymath}
\begin{equation}
\mathbf{A}_{N}^{(k)}=
\left[
\begin{array}{cc}
\mathbf{A}_{N/2}^{(k)}&-\mathbf{A}_{N/2}^{(k-1)}\\
\mathbf{A}_{N/2}^{(k-1)}&\mathbf{A}_{N/2}^{(k)}\\
\end{array} \right] \label{eq:defAN}
\end{equation}
\begin{displaymath}
\mathbf{A}_{N}^{(n)}=
\left[
\begin{array}{cc}
\mathbf{0}_{N/2}&-\mathbf{A}_{N/2}^{(n-1)}\\
\mathbf{A}_{N/2}^{(n-1)}&\mathbf{0}_{N/2}\\
\end{array} \right] 
\end{displaymath}
for $k=1,2,\ldots,n-1$ and $n$=log$_2N$, where
\begin{displaymath}
\mathbf{A}_{2}^{(0)}=
\left[
\begin{array}{cc}
1&0\\
0&1\\
\end{array} \right]=\mathbf{I}_{2},\;\;\;
\mathbf{A}_{2}^{(1)}=
\left[
\begin{array}{cc}
0&-1\\
1&0\\
\end{array} \right].
\end{displaymath}
To clarify our idea we show the explicit form of expressions (\ref{eq:sum}) and (\ref{eq:defAN}) for $N=2$, $N=4$ and $N=8$ in an Example \ref{Example3}.

\begin{figure}[h!]
\centering
\includegraphics[width=1.0\textwidth]{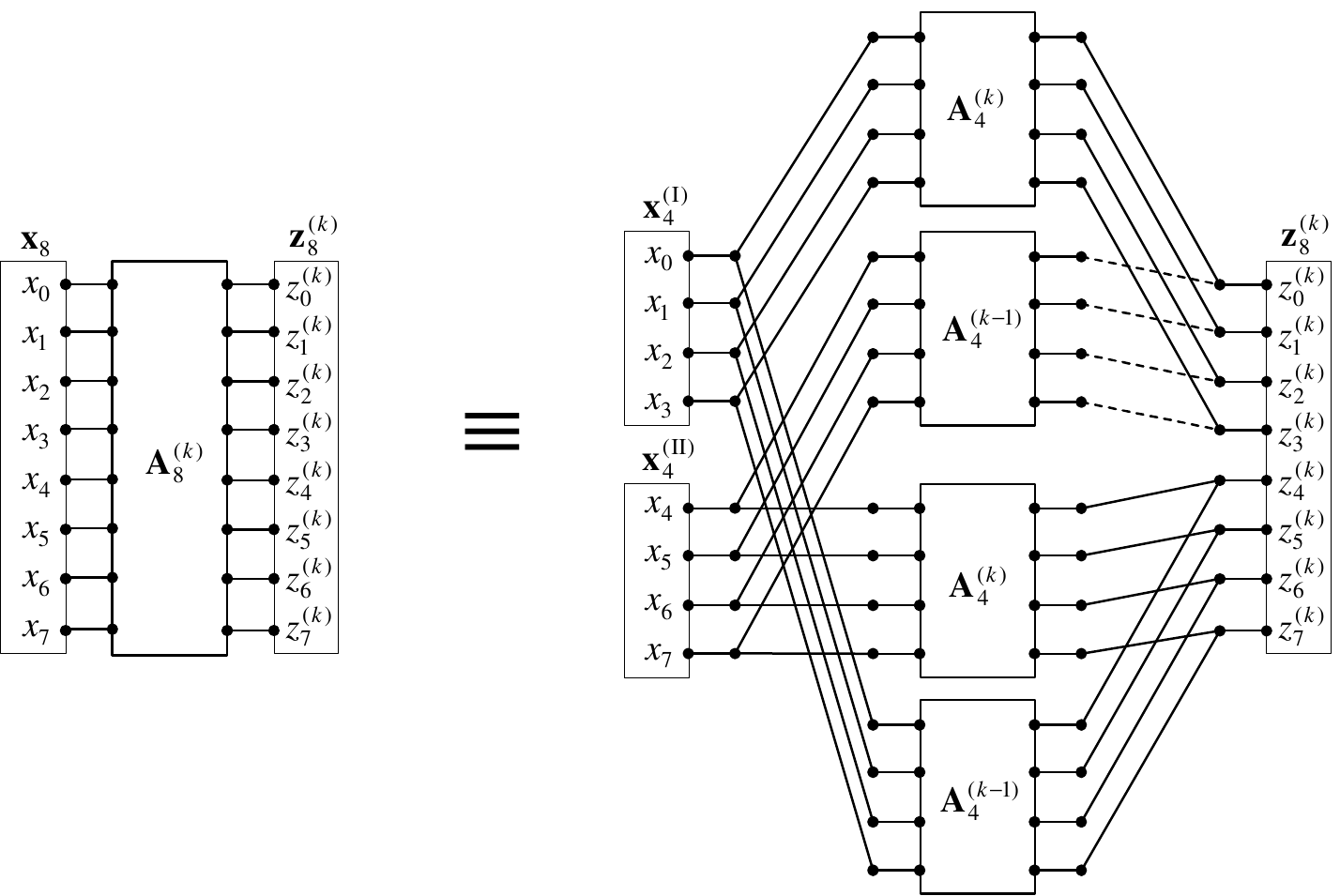} 
\caption{The way of calculating the products $\mathbf{A}_8^{(k)}\mathbf{x}_8$ using the products of twice smaller size:  $\mathbf{A}_{4}^{(k)}\mathbf{x}_{4}^{(I)}$, $\mathbf{A}_{4}^{(k-1)}\mathbf{x}_{4}^{(II)}$, $\mathbf{A}_{4}^{(k-1)}\mathbf{x}_{4}^{(I)}$, $\mathbf{A}_{4}^{(k)}\mathbf{x}_{4}^{(II)}$ for $k=1,2$}
\label{fig:figure2}  
\end{figure}

\begin{example} \label{Example3}
\begin{displaymath}
\mathbf{\overline V}_2=\mathbf{A}_2^{(0)}+b\mathbf{A}_2^{(1)}
\end{displaymath}
where the matrices $\mathbf{A}_2^{(0)}$ and $\mathbf{A}_2^{(1)}$ are presented above.
\begin{displaymath}
\mathbf{\overline V}_4=\mathbf{A}_4^{(0)}+b\mathbf{A}_4^{(1)}+b^2\mathbf{A}_4^{(2)}
\end{displaymath}
where
\begin{displaymath}
\mathbf{A}_4^{(0)}=
\left[
\begin{array}{cccc}
1&0&0&0\\
0&1&0&0\\
0&0&1&0\\
0&0&0&1\\
\end{array} \right]=
\left[
\begin{array}{cc}
\mathbf{A}_{2}^{(0)}&\mathbf{0}_{2}\\
\mathbf{0}_{2}&\mathbf{A}_{2}^{(0)}\\
\end{array} \right]=
\mathbf{I}_4,
\end{displaymath}

\begin{displaymath}
\mathbf{A}_4^{(1)}=
\left[
\begin{array}{cccc}
0&-1&-1&0\\
1&0&0&-1\\
1&0&0&-1\\
0&1&1&0\\
\end{array} \right]=
\left[
\begin{array}{cc}
\mathbf{A}_{2}^{(1)}&-\mathbf{A}_{2}^{(0)}\\
\mathbf{A}_{2}^{(0)}&\mathbf{A}_{2}^{(1)}\\
\end{array} \right],
\end{displaymath}

\begin{displaymath}
\mathbf{A}_4^{(2)}=
\left[
\begin{array}{cccc}
0&0&0&1\\
0&0&-1&0\\
0&-1&0&0\\
1&0&0&0\\
\end{array} \right]=
\left[
\begin{array}{cc}
\mathbf{0}_{2}&-\mathbf{A}_{2}^{(1)}\\
\mathbf{A}_{2}^{(1)}&\mathbf{0}_{2}\\
\end{array} \right] .
\end{displaymath}

\begin{displaymath}
\mathbf{\overline V}_8=\mathbf{A}_8^{(0)}+b\mathbf{A}_8^{(1)}+b^2\mathbf{A}_8^{(2)}+b^3\mathbf{A}_8^{(3)},
\end{displaymath}
where
\begin{displaymath}
\mathbf{A}_8^{(0)}=
\left[
\begin{array}{cccccccc}
1&0&0&0&0&0&0&0\\
0&1&0&0&0&0&0&0\\
0&0&1&0&0&0&0&0\\
0&0&0&1&0&0&0&0\\
0&0&0&0&1&0&0&0\\
0&0&0&0&0&1&0&0\\
0&0&0&0&0&0&1&0\\
0&0&0&0&0&0&0&1\\
\end{array} \right]=
\left[
\begin{array}{cc}
\mathbf{A}_{4}^{(0)}&\mathbf{0}_{4}\\
\mathbf{0}_{4}&\mathbf{A}_{4}^{(0)}\\
\end{array} \right]=
\mathbf{I}_8,
\end{displaymath}

\begin{displaymath}
\mathbf{A}_8^{(1)}=
\left[
\begin{array}{cccccccc}
0&-1&-1&0&-1&0&0&0\\
1&0&0&-1&0&-1&0&0\\
1&0&0&-1&0&0&-1&0\\
0&1&1&0&0&0&0&-1\\
1&0&0&0&0&-1&-1&0\\
0&1&0&0&1&0&0&-1\\
0&0&1&0&1&0&0&-1\\
0&0&0&1&0&1&1&0\\
\end{array} \right]=
\left[
\begin{array}{cc}
\mathbf{A}_{4}^{(1)}&-\mathbf{A}_{4}^{(0)}\\
\mathbf{A}_{4}^{(0)}&\mathbf{A}_{4}^{(1)}\\
\end{array} \right],
\end{displaymath}

\begin{displaymath}
\mathbf{A}_8^{(2)}=
\left[
\begin{array}{cccccccc}
0&0&0&1&0&1&1&0\\
0&0&-1&0&-1&0&0&1\\
0&-1&0&0&-1&0&0&1\\
1&0&0&0&0&-1&-1&0\\
0&-1&-1&0&0&0&0&1\\
1&0&0&-1&0&0&-1&0\\
1&0&0&-1&0&-1&0&0\\
0&1&1&0&1&0&0&0\\
\end{array} \right]=
\left[
\begin{array}{cc}
\mathbf{A}_{4}^{(2)}&-\mathbf{A}_{4}^{(1)}\\
\mathbf{A}_{4}^{(1)}&\mathbf{A}_{4}^{(2)}\\
\end{array} \right],
\end{displaymath}

\begin{displaymath}
\mathbf{A}_8^{(3)}=
\left[
\begin{array}{cccccccc}
0&0&0&0&0&0&0&-1\\
0&0&0&0&0&0&1&0\\
0&0&0&0&0&1&0&0\\
0&0&0&0&-1&0&0&0\\
0&0&0&1&0&0&0&0\\
0&0&-1&0&0&0&0&0\\
0&-1&0&0&0&0&0&0\\
1&0&0&0&0&0&0&0\\
\end{array} \right]=
\left[
\begin{array}{cc}
\mathbf{0}_{4}&-\mathbf{A}_{4}^{(2)}\\
\mathbf{A}_{4}^{(2)}&\mathbf{0}_{4}\\
\end{array} \right].
\end{displaymath}
\end{example}
Now we will evaluate the number of arithmetical operations, which are necessary to calculate the matrix-vector product $\mathbf{y}_N=\mathbf{\overline{V}}_N\mathbf{x}_N$. We note, that in a general case such an operation requires $N^2$ multiplications and  $N(N-1)$ additions.
Now we will calculate the numbers of multiplications and additions needed for this operation if we use the expression (\ref{eq:sum}) for the matrix  $\mathbf{\overline{V}}_N$, i.e.
\begin{equation}
\mathbf{y}_N=\mathbf{A}_N^{(0)}\mathbf{x}_N+b\mathbf{A}_N^{(1)}\mathbf{x}_N+b^2\mathbf{A}_N^{(2)}\mathbf{x}_N+\ldots
+b^n\mathbf{A}_N^{(n)}\mathbf{x}_N \label{eq:rozklad4}
\end{equation}
Since the non-zero entries of matrices $\mathbf{A}_N^{(0)}$,  $\mathbf{A}_N^{(1)}, \ldots, \mathbf{A}_N^{(n)}$ are only 1 and -1, no multiplications are needed when calculating the matrix-vector products $\mathbf{A}_{N}^{(k)}\mathbf{x}_N$. The only multiplications we have to perform are multiplications of the vectors $\mathbf{A}_N^{(k)}\mathbf{x}_N$ by the powers of $b$: $\mathbf{A}_N^{(1)}\mathbf{x}_N$ by $b$, $\mathbf{A}_N^{(2)}\mathbf{x}_N$ by $b^2,\ldots, \mathbf{A}_N^{(n)}\mathbf{x}_N$ by $b^n$. Because the number $b$ is constant and known, its powers $b^2$, $b^3,\ldots,b^n$ may be prepared in advance. Thus, the number of multiplication by calculating the matrix-vector product $\mathbf{\overline{V}}_N\mathbf{x}_N$ is equal to $nN=N$log$N$.
Let us examine the number of additions, we need to perform, when calculating the matrix-vector product $\mathbf{\overline{V}}_N\mathbf{x}_N$.
The total number of additions consist of number of additions by calculating the matrix-vector products $\mathbf{A}_N^{(k)}\mathbf{x}_N$,  and $nN$ additions which are needed to calculate the sum of vectors: $\mathbf{A}_N^{(0)}\mathbf{x}_N$, $b\mathbf{A}_N^{(1)}\mathbf{x}_N$, $b^2\mathbf{A}_N^{(2)}\mathbf{x}_N,\ldots,b^n\mathbf{A}_N^{(n)}\mathbf{x}_N$. Since, according to (\ref{eq:defAN}), the matrices $\mathbf{A}_N^{(k)}$  have specific structures, the products $\mathbf{A}_N^{(k)}\mathbf{x}_N$ can be obtained by subtracting the products $\mathbf{A}_{N/2}^{(k)}\mathbf{x}_{N/2}^{(I)}$, $\mathbf{A}_{N/2}^{(k-1)}\mathbf{x}_{N/2}^{(II)}$ of twice smaller size and summing the products $\mathbf{A}_{N/2}^{(k-1)}\mathbf{x}_{N/2}^{(I)}$, $\mathbf{A}_{N/2}^{(k)}\mathbf{x}_{N/2}^{(II)}$ (excluding products $\mathbf{A}_N^{(0)}\mathbf{x}_N$ and $\mathbf{A}_N^{(n)}\mathbf{x}_N$ which can be obtained even in a simpler way). By $\mathbf{x}_{N/2}^{(I)}=[x_0,x_1,\ldots,x_{N/2-1}]^T$ we denote the first half of the input vector $\mathbf{x}_N$ and by $\mathbf{x}_{N/2}^{(II)}=[x_{N/2},x_{N/2+1},\ldots,x_{N-1}]^T$ - the second half of this vector, as it was shown, for $N=8$, in the Figure \ref{fig:figure2}.

\begin{figure}[h!]
\centering
\includegraphics[width=0.7\textwidth]{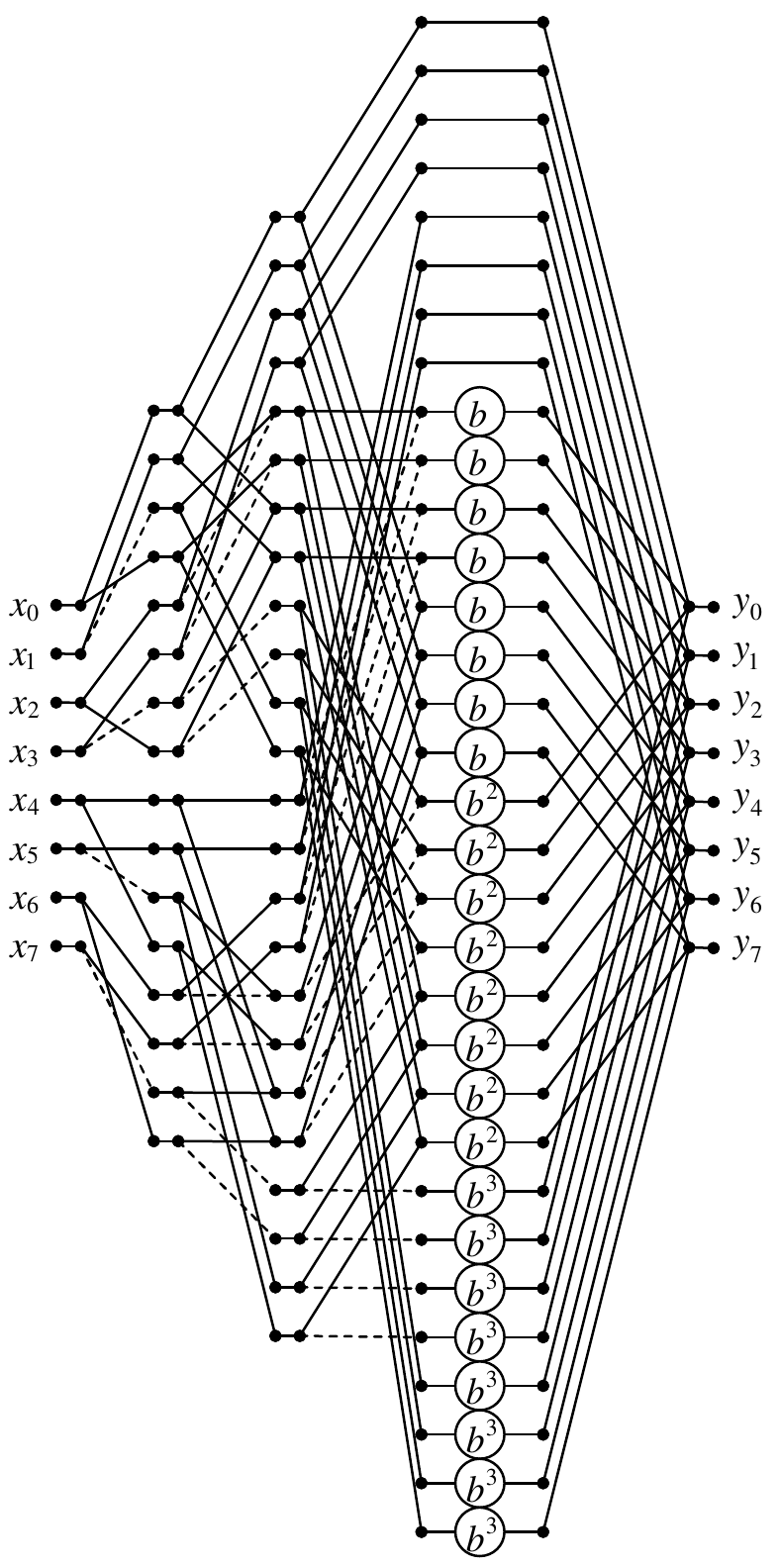} 
\caption{Data flow diagram of calculating the products $\mathbf{y}_8=\mathbf{\overline{V}}_8\mathbf{x}_8$}
\label{fig:figure3}  
\end{figure}

It should be noted that the products $\mathbf{A}_{N/2}^{(k)}\mathbf{x}_{N/2}^{(I)}$ and $\mathbf{A}_{N/2}^{(k)}\mathbf{x}_{N/2}^{(II)}$ are used to calculate both products $\mathbf{A}_{N}^{(k)}\mathbf{x}_{N}$ and $\mathbf{A}_{N}^{(k+1)}\mathbf{x}_{N}$. For example the products $\mathbf{A}_{4}^{(0)}\mathbf{x}_{4}^{(I)}$ and $\mathbf{A}_{4}^{(0)}\mathbf{x}_{4}^{(II)}$ are used to calculate $\mathbf{A}_{8}^{(0)}\mathbf{x}_{8}$ and $\mathbf{A}_{8}^{(1)}\mathbf{x}_{8}$. It allows to reduce the number of additions, because the some products are used twice. 
Of course, this procedure can be repeated and each of products $\mathbf{A}_{N/2}^{(k)}\mathbf{x}_{N/2}^{(I)}$, $\mathbf{A}_{N/2}^{(k-1)}\mathbf{x}_{N/2}^{(II)}$, $\mathbf{A}_{N/2}^{(k-1)}\mathbf{x}_{N/2}^{(I)}$, $\mathbf{A}_{N/2}^{(k)}\mathbf{x}_{N/2}^{(II)}$ can be calculated by summing (subtracting) products of twice smaller size and so on. It can be continued until calculating products of matrices $\mathbf{A}_{2}^{(0)}$ and $\mathbf{A}_{2}^{(1)}$ by two-element sub-vectors of the vector $\mathbf{x}_{N}$. The whole process of going down by calculating the product $\mathbf{y}_N=\mathbf{\overline{V}}_N\mathbf{x}_N$ is presented,  for $N=8$, in the Figure \ref{fig:figure3}.

The expression (\ref{eq:rozklad4}) can be also written as the matrix-vector product, as follows:
\begin{equation}
\mathbf{y}_N=\mathbf{\overline{V}}_N\mathbf{x}_N=\mathbf{C}_{N\times{(n+1)N}}\mathbf{B}_{(n+1)N}\mathbf{A}_{(n+1)N\times{N}}\mathbf{x}_N \label{eq:rozkladmac}
\end{equation}
where
\begin{displaymath}
\mathbf{A}_{(n+1)N\times{N}}=
\left[
\begin{array}{c}
\mathbf{A}_N^{(0)}\\
\mathbf{A}_N^{(1)}\\
\vdots\\
\mathbf{A}_N^{(n)}\\
\end{array} \right],
\end{displaymath}
\begin{displaymath}
\mathbf{B}_{(n+1)N}=
\left[
\begin{array}{cccc}
1&0&\ldots&0\\
0&b&\ldots&0\\
\vdots&\vdots&\ddots&\vdots\\
0&0&\ldots&b^n\\
\end{array} \right]\otimes\mathbf{I}_N,
\end{displaymath}
\begin{displaymath}
\mathbf{C}_{N\times{(n+1)N}}=\mathbf{1}_{1\times{(n+1)}}\otimes\mathbf{I}_N.
\end{displaymath}
The symbol $\otimes$ denotes the Kronecker product of matrices, and $\mathbf{1}_{1\times{(n+1)}}$ is the matrix (row vector) whose all entries are equal to 1.
The matrix $\mathbf{A}_{(n+1)N\times{N}}$ is responsible for multiplications of the matrices $\mathbf{A}_N^{(0)}$, $\mathbf{A}_N^{(1)}$, $\ldots$, $\mathbf{A}_N^{(n)}$ by the input vector $\mathbf{x}_N$, the matrix $\mathbf{B}_{(n+1)N}$ - for multiplications of those products by the proper powers of $b$, and the matrix $\mathbf{C}_{N\times{nN}}$ - for aggregation of results.
The process of going down by calculating the product $\mathbf{\overline{V}}_N\mathbf{x}_N$, which has been presented in Figures \ref{fig:figure2} and \ref{fig:figure3}, can be also described in the therm of matrices product.
It means factorisation of the matrix $\mathbf{A}_{(n+1)N\times{N}}$  into the product of $n$ matrices
\begin{equation}
\mathbf{A}_{(n+1)N\times{N}} =\mathbf{A}_{(n+1)N\times{nN}}\mathbf{A}_{nN\times{(n-1)N}}\ldots\mathbf{A}_{2N\times{N}} \label{eq:rozkladA}
\end{equation}
The matrices which occur on the right side of expression (\ref{eq:rozkladA}) have the following forms:
\begin{equation}
\mathbf{A}_{2N\times{N}}=\mathbf{I}_{N/2}\otimes
\left[
\begin{array}{c}
\mathbf{\overline{A}}_{2\times 2}^{(0)}\\
\mathbf{\overline{A}}_{2\times 2}^{(1)}\\
\end{array} \right] 
\end{equation}
where
\begin{displaymath}
\mathbf{\overline{A}}_{2\times 2}^{(0)}=\mathbf{A}_{2}^{(0)}\otimes\left[ 1\right]\otimes\mathbf{I}_{1}=\mathbf{A}_{2}^{(0)},\;\;
\mathbf{\overline{A}}_{2\times 2}^{(1)}=\mathbf{A}_{2}^{(1)}\otimes\left[ 1\right]\otimes\mathbf{I}_{1}=\mathbf{A}_{2}^{(1)}.
\end{displaymath}
\begin{equation}
\mathbf{A}_{3N\times{2N}} =\mathbf{I}_{N/4}\otimes
\left[
\begin{array}{c}
\mathbf{\overline{A}}_{4\times 8}^{(0)}\\
\mathbf{\overline{A}}_{4\times 8}^{(0)(2\rightarrow)}+\mathbf{\overline{A}}_{4\times 8}^{(1)(2\leftarrow)}\\
\mathbf{\overline{A}}_{4\times 8}^{(1)}\\
\end{array} \right], 
\end{equation}
where
\begin{displaymath}
\mathbf{\overline{A}}_{4\times 8}^{(0)}=\mathbf{A}_{2}^{(0)}\otimes\left[ 1\;\;0\right]\otimes\mathbf{I}_{2},\;\;
\mathbf{\overline{A}}_{4\times 8}^{(1)}=\mathbf{A}_{2}^{(1)}\otimes\left[0\;\; 1\right]\otimes\mathbf{I}_{2}
\end{displaymath}
and the matrix $\mathbf{\overline{A}}_{4\times 8}^{(1)(2\rightarrow)}$ denotes the matrix $\mathbf{\overline{A}}_{4\times 8}^{(1)}$ which columns  were circularly shifted by 2 positions to the right, and  the matrix $\mathbf{\overline{A}}_{4\times 8}^{(1)(2\leftarrow)}$ denotes the matrix $\mathbf{\overline{A}}_{4\times 8}^{(1)}$ which columns were circularly shifted by 2 positions to the left.
The last matrix $\mathbf{A}_{(n+1)N\times{nN}}$ is defined as
\begin{equation}
\mathbf{A}_{(n+1)N\times{nN}} =\mathbf{I}_{N/N}\otimes
\left[
\begin{array}{c}
\mathbf{\overline{A}}_{N\times nN}^{(0)}\\
\mathbf{\overline{A}}_{N\times nN}^{(0)(N/2\rightarrow)}+\mathbf{\overline{A}}_{N\times nN}^{(1)((n-1)N/2\leftarrow)}\\
\vdots\\
\mathbf{\overline{A}}_{N\times nN}^{(0)((n-1)N/2\rightarrow)}+\mathbf{\overline{A}}_{N\times nN}^{(1)(N/2\leftarrow)}\\
\mathbf{\overline{A}}_{N\times nN}^{(1)}\\
\end{array} \right] 
\end{equation}
where
\begin{displaymath}
\mathbf{\overline{A}}_{N\times nN}^{(0)}=\mathbf{A}_{2}^{(0)}\otimes\left[ 1\;\;0\;\;\ldots\;\;0\right]\otimes\mathbf{I}_{N/2},\;\;
\mathbf{\overline{A}}_{N\times nN}^{(1)}=\mathbf{A}_{2}^{(1)}\otimes\left[0\;\;0\;\;\ldots \;\;1\right]\otimes\mathbf{I}_{N/2}.
\end{displaymath}
Using the expression (\ref{eq:rozkladA}) the algorithm (\ref{eq:rozkladmac}) of calculating the product $\mathbf{\overline{V}}_N\mathbf{x}_N$  can be written as follows:
\begin{equation}
\mathbf{y}_N=\mathbf{C}_{N\times{(n+1)N}}\mathbf{B}_{(n+1)N}\mathbf{A}_{(n+1)N\times{nN}}\mathbf{A}_{nN\times{(n-1)N}}\ldots\mathbf{A}_{2N\times{N}}\mathbf{x}_N \label{eq:productV}
\end{equation}
The expression (\ref{eq:productV}) allows for evaluating the total number of additions, which are needed to calculate the matrix-vector product $\mathbf{\overline{V}}_N\mathbf{x}_N$. We assume that the input vector $\mathbf{x}_N$ is real-valued. Each of matrices $\mathbf{A}_{(k+1)N\times{kN}}$, for $k=1,2,\ldots,n$, is the direct sum of $N/2^k$ identical blocks and the single block is the vertical concatenation of $k+1$ matrices. The firs, indicated by $\mathbf{\overline{A}}_{2^k\times k2^k}^{(0)}$, and the last, indicated by $\mathbf{\overline{A}}_{2^k\times k2^k}^{(1)}$, do not need any additions or subtractions by multiplying them by a vector. The $k-1$ others matrices, which are sums of $\mathbf{\overline{A}}_{2^k\times k2^k}^{(0)}$ and $\mathbf{\overline{A}}_{2^k\times k2^k}^{(1)}$, after circularly shifting their columns, so multiplying each of them by a vector needs $2^k$ additions. To calculate the product  $\mathbf{A}_{(n+1)N\times{nN}}\mathbf{A}_{nN\times{(n-1)N}}\ldots\mathbf{A}_{2N\times{N}}\mathbf{x}_N $ we have to perform $\sum_{k=1}^n \frac{N}{2^k}(k-1)2^k=Nn(n-1)/2$ additions. The product of the matrix $\mathbf{B}_{(n+1)N}$ by a vector do not need any additions and the product of the matrix
$\mathbf{C}_{N\times{(n+1)N}}$ by a vector needs $nN$ additions. Thus the total number of additions by calculating the products $\mathbf{\overline{V}}_N\mathbf{x}_N$, according to (\ref{eq:productV}), is equal to $Nn(n+1)/2$.

Example \ref{Example4} shows the explicit form of the algorithm (\ref{eq:productV}) with all occurring in it matrices for $N=8$. 

\begin{example}\label{Example4}
The algorithm (\ref{eq:productV}) of calculating the product of matrix $\mathbf{\overline{V}}_8$ by the vector $\mathbf{x}_8$ is as follows:
\begin{displaymath}
\mathbf{y}_8=\mathbf{\overline{V}}_8\mathbf{x}_8=\mathbf{C}_{8\times{32}}\mathbf{B}_{32}\mathbf{A}_{32\times{24}}\mathbf{A}_{24\times{16}}\mathbf{A}_{16\times{8}}\mathbf{x}_8, 
\end{displaymath}
where
\begin{displaymath}
\mathbf{A}_{16\times{8}}=\mathbf{I}_{4}\otimes
\left[
\begin{array}{c}
\mathbf{\overline{A}}_{2\times 2}^{(0)}\\
\mathbf{\overline{A}}_{2\times 2}^{(1)}\\
\end{array} \right]=
 \left[
 \begin{array}{cccccccc}
 1&0&0&0&0&0&0&0\\
 0&1&0&0&0&0&0&0\\
 0&-1&0&0&0&0&0&0\\
 1&0&0&0&0&0&0&0\\
 0&0&1&0&0&0&0&0\\
 0&0&0&1&0&0&0&0\\
 0&0&0&-1&0&0&0&0\\
 0&0&1&0&0&0&0&0\\
 0&0&0&0&1&0&0&0\\
 0&0&0&0&0&1&0&0\\
 0&0&0&0&0&-1&0&0\\
 0&0&0&0&1&0&0&0\\
 0&0&0&0&0&0&1&0\\
 0&0&0&0&0&0&0&1\\
 0&0&0&0&0&0&0&-1\\
 0&0&0&0&0&0&1&0\\
 \end{array} \right],
 \end{displaymath}
 \begin{displaymath}
 \mathbf{A}_{24\times{16}}=\mathbf{I}_{2}\otimes
 \left[
 \begin{array}{c}
 \mathbf{\overline{A}}_{4\times 8}^{(0)}\\
 \mathbf{\overline{A}}_{4\times 8}^{(0)(2\rightarrow)}+\mathbf{\overline{A}}_{4\times 8}^{(1)(2\leftarrow)}\\
 \mathbf{\overline{A}}_{4\times 8}^{(1)}\\
 \end{array} \right]=
 \end{displaymath}
  \begin{displaymath}
  \left[
  \begin{array}{cccccccccccccccc}
  1&0&0&0&0&0&0&0&0&0&0&0&0&0&0&0\\
  0&1&0&0&0&0&0&0&0&0&0&0&0&0&0&0\\
  0&0&0&0&1&0&0&0&0&0&0&0&0&0&0&0\\
  0&0&0&0&0&1&0&0&0&0&0&0&0&0&0&0\\
  0&0&1&0&-1&0&0&0&0&0&0&0&0&0&0&0\\
  0&0&0&1&0&-1&0&0&0&0&0&0&0&0&0&0\\
  1&0&0&0&0&0&1&0&0&0&0&0&0&0&0&0\\
  0&1&0&0&0&0&0&1&0&0&0&0&0&0&0&0\\
  0&0&0&0&0&0&-1&0&0&0&0&0&0&0&0&0\\
  0&0&0&0&0&0&0&-1&0&0&0&0&0&0&0&0\\
  0&0&1&0&0&0&0&0&0&0&0&0&0&0&0&0\\
  0&0&0&1&0&0&0&0&0&0&0&0&0&0&0&0\\
  0&0&0&0&0&0&0&0&1&0&0&0&0&0&0&0\\
  0&0&0&0&0&0&0&0&0&1&0&0&0&0&0&0\\
  0&0&0&0&0&0&0&0&0&0&0&0&1&0&0&0\\
  0&0&0&0&0&0&0&0&0&0&0&0&0&1&0&0\\
  0&0&0&0&0&0&0&0&0&0&1&0&-1&0&0&0\\
  0&0&0&0&0&0&0&0&0&0&0&1&0&-1&0&0\\
  0&0&0&0&0&0&0&0&1&0&0&0&0&0&1&0\\
  0&0&0&0&0&0&0&0&0&1&0&0&0&0&0&1\\
  0&0&0&0&0&0&0&0&0&0&0&0&0&0&-1&0\\
  0&0&0&0&0&0&0&0&0&0&0&0&0&0&0&-1\\
  0&0&0&0&0&0&0&0&0&0&1&0&0&0&0&0\\
  0&0&0&0&0&0&0&0&0&0&0&1&0&0&0&0\\
  \end{array} \right],
 \end{displaymath}
\begin{displaymath}
 \mathbf{A}_{32\times{24}}=\mathbf{I}_{1}\otimes
 \left[
 \begin{array}{c}
 \mathbf{\overline{A}}_{8\times 24}^{(0)}\\
 \mathbf{\overline{A}}_{8\times 24}^{(0)(4\rightarrow)}+\mathbf{\overline{A}}_{8\times 8}^{(1)(8\leftarrow)}\\
 \mathbf{\overline{A}}_{8\times 24}^{(0)(8\rightarrow)}+\mathbf{\overline{A}}_{8\times 8}^{(1)(4\leftarrow)}\\
 \mathbf{\overline{A}}_{8\times 24}^{(1)}\\
 \end{array} \right]=
\end{displaymath}
  \begin{displaymath}
  \left[
{\setlength\arraycolsep{2pt}
\begin{array}{cccccccccccccccccccccccc}
1&0&0&0&0&0&0&0&0&0&0&0&0&0&0&0&0&0&0&0&0&0&0&0\\
0&1&0&0&0&0&0&0&0&0&0&0&0&0&0&0&0&0&0&0&0&0&0&0\\
0&0&1&0&0&0&0&0&0&0&0&0&0&0&0&0&0&0&0&0&0&0&0&0\\
0&0&0&1&0&0&0&0&0&0&0&0&0&0&0&0&0&0&0&0&0&0&0&0\\
0&0&0&0&0&0&0&0&0&0&0&0&1&0&0&0&0&0&0&0&0&0&0&0\\
0&0&0&0&0&0&0&0&0&0&0&0&0&1&0&0&0&0&0&0&0&0&0&0\\
0&0&0&0&0&0&0&0&0&0&0&0&0&0&1&0&0&0&0&0&0&0&0&0\\
0&0&0&0&0&0&0&0&0&0&0&0&0&0&0&1&0&0&0&0&0&0&0&0\\
0&0&0&0&1&0&0&0&0&0&0&0&-1&0&0&0&0&0&0&0&0&0&0&0\\
0&0&0&0&0&1&0&0&0&0&0&0&0&-1&0&0&0&0&0&0&0&0&0&0\\
0&0&0&0&0&0&1&0&0&0&0&0&0&0&-1&0&0&0&0&0&0&0&0&0\\
0&0&0&0&0&0&0&1&0&0&0&0&0&0&0&-1&0&0&0&0&0&0&0&0\\
1&0&0&0&0&0&0&0&0&0&0&0&0&0&0&0&1&0&0&0&0&0&0&0\\
0&1&0&0&0&0&0&0&0&0&0&0&0&0&0&0&0&1&0&0&0&0&0&0\\
0&0&1&0&0&0&0&0&0&0&0&0&0&0&0&0&0&0&1&0&0&0&0&0\\
0&0&0&1&0&0&0&0&0&0&0&0&0&0&0&0&0&0&0&1&0&0&0&0\\
0&0&0&0&0&0&0&0&1&0&0&0&0&0&0&0&-1&0&0&0&0&0&0&0\\
0&0&0&0&0&0&0&0&0&1&0&0&0&0&0&0&0&-1&0&0&0&0&0&0\\
0&0&0&0&0&0&0&0&0&0&1&0&0&0&0&0&0&0&-1&0&0&0&0&0\\
0&0&0&0&0&0&0&0&0&0&0&1&0&0&0&0&0&0&0&-1&0&0&0&0\\
0&0&0&0&1&0&0&0&0&0&0&0&0&0&0&0&0&0&0&0&1&0&0&0\\
0&0&0&0&0&1&0&0&0&0&0&0&0&0&0&0&0&0&0&0&0&1&0&0\\
0&0&0&0&0&0&1&0&0&0&0&0&0&0&0&0&0&0&0&0&0&0&1&0\\
0&0&0&0&0&0&0&1&0&0&0&0&0&0&0&0&0&0&0&0&0&0&0&1\\
0&0&0&0&0&0&0&0&0&0&0&0&0&0&0&0&0&0&0&0&-1&0&0&0\\
0&0&0&0&0&0&0&0&0&0&0&0&0&0&0&0&0&0&0&0&0&-1&0&0\\
0&0&0&0&0&0&0&0&0&0&0&0&0&0&0&0&0&0&0&0&0&0&-1&0\\
0&0&0&0&0&0&0&0&0&0&0&0&0&0&0&0&0&0&0&0&0&0&0&-1\\
0&0&0&0&0&0&0&0&1&0&0&0&0&0&0&0&0&0&0&0&0&0&0&0\\
0&0&0&0&0&0&0&0&0&1&0&0&0&0&0&0&0&0&0&0&0&0&0&0\\
0&0&0&0&0&0&0&0&0&0&1&0&0&0&0&0&0&0&0&0&0&0&0&0\\
0&0&0&0&0&0&0&0&0&0&0&1&0&0&0&0&0&0&0&0&0&0&0&0\\
\end{array}} \right],
\end{displaymath}
\begin{displaymath}
 \mathbf{B}_{32}=
 \left[
 \begin{array}{cccc}
 1&0&0&0\\
 0&b&0&0\\
 0&0&b^2&0\\
 0&0&0&b^3\\
 \end{array} \right]\otimes \mathbf{I}_{8},
 \end{displaymath}
\begin{displaymath}
\mathbf{C}_{8\times{32}}=\mathbf{1}_{1\times 4}\otimes \mathbf{I}_{8}.
\end{displaymath}
It is easy to check that in this case the total number of addition is equal to 48 and the number of multiplications is equal to 24 (we can see it also in the figure \ref{fig:figure3}).
\end{example}

\section{The novel DFRHT algorithm}
\begin{figure}[h]
\centering
\includegraphics[width=1.0\textwidth]{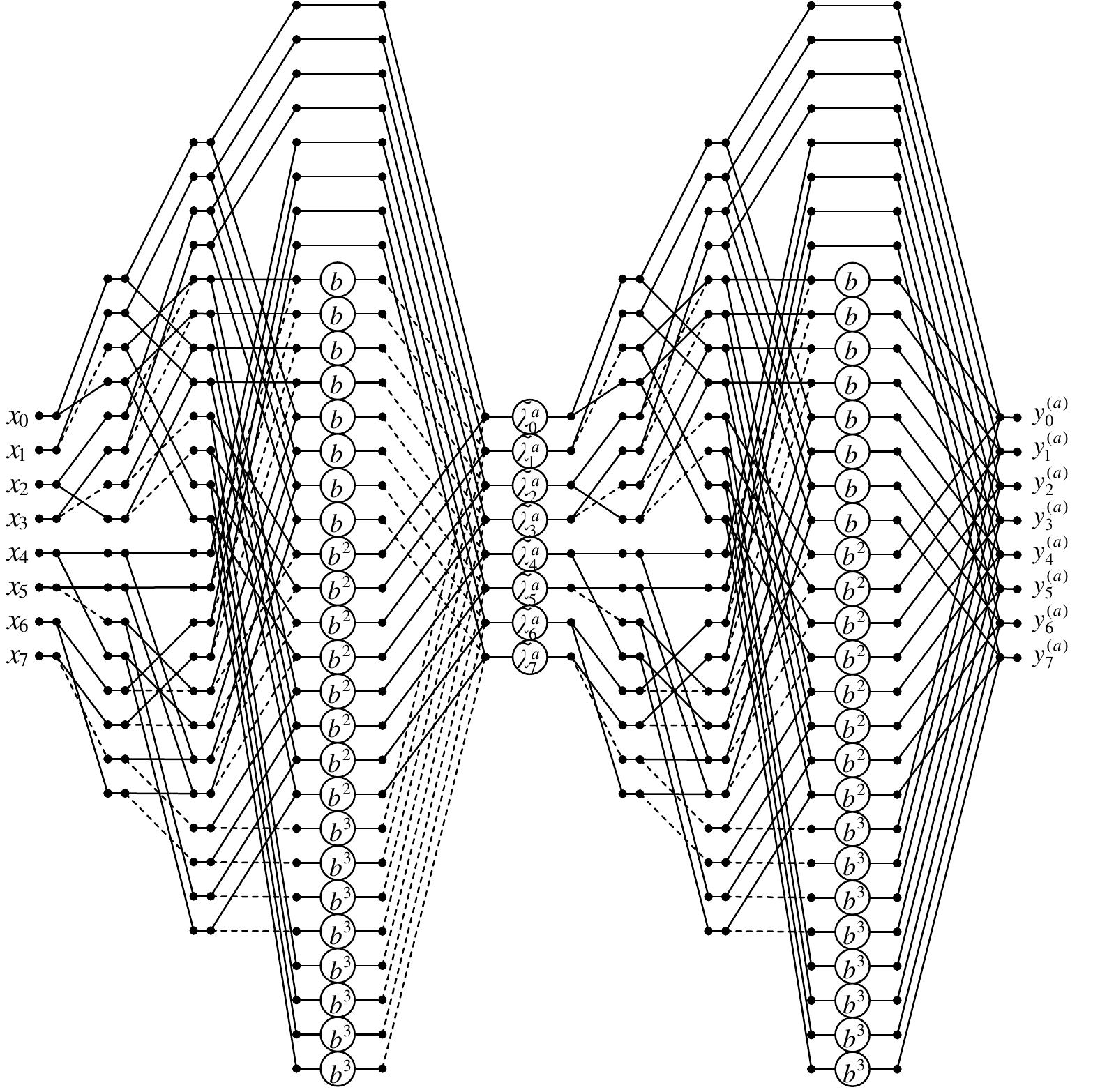} 
\caption{Data flow diagram of the DFRHT algorithm (\ref{eq:transformata2}) for $N=8$}
\label{fig:figure4}  
\end{figure}

Now we return to the DFRHT algorithm (\ref{eq:transformatan}). According to (\ref{eq:sum}) the matrix $\mathbf{\overline V}_N$ can be written as the sum of the matrices  $\mathbf{A}_N^{(0)}$,  $\mathbf{A}_N^{(1)}, \ldots, \mathbf{A}_N^{(n)}$ with coefficients $1, b, \ldots, b^n$. The transposed matrix $\mathbf{\overline V}_N^T$ can be  written as the sum of the transposed matrices  $\mathbf{A}_N^{(0)T}$,  $\mathbf{A}_N^{(1)T}, \ldots, \mathbf{A}_N^{(n)T}$ with the same coefficients $1, b, \ldots, b^n$:
\begin{equation}
\mathbf{\overline V}_N^T=\mathbf{A}_N^{(0)T}+b\mathbf{A}_N^{(1)T}+b^2\mathbf{A}_N^{(2)T}+\ldots+b^n\mathbf{A}_N^{(n)T} \label{eq:sumT}
\end{equation}
Since the matrices  with the even indexes $\mathbf{A}_N^{(0)}$,  $\mathbf{A}_N^{(2)}, \ldots$ are symmetric and the matrices with the odd indexes are asymmetric the expression (\ref{eq:sumT}) can be written as follows:
\begin{equation}
\mathbf{\overline V}_N^T=\mathbf{A}_N^{(0)}-b\mathbf{A}_N^{(1)}+\ldots+(-1)^n b^n\mathbf{A}_N^{(n)} \label{eq:sumT1}
\end{equation}
According to (\ref{eq:rozkladmac}) the matrix $\mathbf{\overline V}_N$ from expression (\ref{eq:sum}) can be transformed into the product 
\begin{equation}
\mathbf{\overline{V}}_N=\mathbf{C}_{N\times{(n+1)N}}\mathbf{B}_{(n+1)N}\mathbf{A}_{(n+1)N\times{N}} \label{eq:rozkladmac1}
\end{equation}
so the matrix $\mathbf{\overline V}_N^T$ may by also transformed from (\ref{eq:sumT1}) into the following product:
\begin{equation}
\mathbf{\overline{V}}_N^T=\mathbf{\overline C}_{N\times{(n+1)N}}\mathbf{B}_{(n+1)N}\mathbf{A}_{(n+1)N\times{N}} \label{eq:rozkladmac1T}
\end{equation}
where the matrix $\mathbf{\overline C}_{N\times{(n+1)N}}$ is defined as follows:
\begin{displaymath}
\mathbf{\overline C}_{N\times{(n+1)N}}=\mathbf{\overline 1}_{1\times{(n+1)}}\otimes\mathbf{I}_N 
\end{displaymath}
and
\begin{displaymath}
\mathbf{\overline 1}_{1\times{(n+1)}}=[1 \hspace{0.2cm} -1 \hspace{0.2cm} \ldots \hspace{0.2cm} (-1)^n].
\end{displaymath}
The others matrices in the expression (\ref{eq:rozkladmac1T}) are the same as that in the expression (\ref{eq:rozkladmac}).
Taking into account the decompositions (\ref{eq:rozkladmac1}) and  (\ref{eq:rozkladmac1T})  of matrices $\mathbf{\overline{V}}_N$ and $\mathbf{\overline{V}}_N^T$ respectively the DFRHT algorithm (\ref{eq:transformatan}) will take the form:
\begin{equation}
\mathbf{y}_{N}^{(a)}=
\mathbf{C}_{N\times{(n+1)N}}\mathbf{B}_{(n+1)N}\mathbf{A}_{(n+1)N\times{N}} \mathbf{\tilde{\Lambda}}_{N}^{a} 
\mathbf{\overline C}_{N\times{(n+1)N}}\mathbf{B}_{(n+1)N}\mathbf{A}_{(n+1)N\times{N}} \mathbf{x}_{N}  \label{eq:transformata2}
\end{equation}
where the matrix $\mathbf{A}_{(n+1)N\times{N}}$ is decomposed according to (\ref{eq:rozkladA}).
For example, for $N=8$  this algorithm  will take the following form:
\begin{displaymath}
\mathbf{y}_{8}^{(a)}=
\mathbf{C}_{8\times{32}}\mathbf{B}_{32}\mathbf{A}_{32\times{24}}\mathbf{A}_{24\times{16}}\mathbf{A}_{16\times{8}} \mathbf{\tilde{\Lambda}}_{8}^{a} 
\mathbf{\overline C}_{8\times {32}}\mathbf{B}_{32}\mathbf{A}_{32\times{24}}\mathbf{A}_{24\times{16}}\mathbf{A}_{16\times{8}} \mathbf{x}_{8}.  \label{eq:transformataN8}
\end{displaymath}
Figure \ref{fig:figure4} shows a data flow diagram of the algorithm for 8 point DFRHT.

\section{Discussion of computational complexity}
Let us assess the computational complexity in term of numbers of multiplications and additions required for DFRHT calculation. Calculation of the discrete fractional Hadamard transform for a real-valued vector $\mathbf{x}_{N}$ of length $N=2^n$, assuming that the matrix $\mathbf{H}_{N}^{a}$ defined by (\ref{eq:rozklad2}) is given, requires $N^2=2^{2n}$ multiplications of a complex number by a real number and $N(N-1)=2^n(2^n-1)$ complex additions. Each multiplication of a complex number by a real number needs two real multiplications and  each addition of two complex numbers requires two real additions. Hence the numbers of real multiplications  and real additions required for computing the DFRHT using the naive method are equal to $2^{2n+1}$ and $2^{n+1}(2^n-1)$ respectively.

Let us now evaluate the computational complexity of the DFRHT realization with the help of the procedure (\ref{eq:transformata2}). As it was discussed in the section \ref{advantages}, if we use the factorized representation of the matrices $\mathbf{\overline{V}}_N^T$ and $\mathbf{\overline{V}}_N$, calculating the product of the real-valued matrix $\mathbf{\overline{V}}_N^T$ and the real-valued vector $\mathbf{x}_{N}$ requires $nN$ real multiplications and $Nn(n+1)/2$ real additions. As a result, we again obtain the real-valued vector.  Then there is computed the product of the complex-valued diagonal matrix $\mathbf{\tilde{\Lambda}}_N^a$ and the real-valued vector calculated previously (we assume that for a predetermined parameter $a$, the diagonal elements of this matrix were calculated in advance). The calculation of this product requires $2N$ real multiplications. The resulting complex-valued vector is then multiplied by the factorized matrix $\mathbf{\overline{V}}_N$. This operation requires  $2Nn$ real multiplications and $Nn(n+1)$ real additions. The total numbers of arithmetic operations to compute DFRHT of size $2^n$ using our new algorithm are $N(3n+2)$ real multiplications and $3Nn(n+1)/2$ real additions. It is easy to check that even for small $n$ the numbers of arithmetic operations required for realization of the proposed algorithm are several times less than in the naive method of computing.

Tables \ref{tab:1} and \ref{tab:2} display the numbers of multiplications and additions required for the DFRHT transform implementation of the real-valued input signal of the length $N=2^n$. These numbers were calculated for three methods of the transform implementation: the direct multiplication of the DFRHT matrix  by a vector of the input data, calculation according to authors' algorithm described in the work \cite{ma}, and according to the algorithm (\ref{eq:transformata2}) proposed in this article.
It is easy to check that  for  $n>5$ the number of arithmetic operations, required for DFRHT transform realization according to the proposed algorithm, is smaller than in the other two methods of DFRHT computing.

\begin{table} [!h]
\caption{Numbers of multiplications for mentioned algorithms}
\label{tab:1}       
\begin{tabular}{rrrr}
\hline\noalign{\smallskip}
$N=2^n$ & direct method & method \cite{ma} & proposed algorithm \\
\noalign{\smallskip}\hline\noalign{\smallskip}
2 & 8 & 6 & 10 \\
4 & 32 & 18 & 32 \\
8 & 128 & 54 & 88 \\
16 & 512 & 162 & 224 \\
32 & 2048 & 486 & 544 \\
64 & 8192 & 1458 & 1280 \\
128 & 32768 & 4374 & 2944 \\
256 & 131072 & 13122 & 6656 \\
512 & 524288 & 39366 & 14848 \\
1024 & 2097152 & 118098 & 32768 \\
\noalign{\smallskip}\hline
\end{tabular}
\end{table}

\begin{table} [!h]
\caption{Numbers of additions for mentioned algorithms}
\label{tab:2}       
\begin{tabular}{rrrr}
\hline\noalign{\smallskip}
$N=2^n$ & direct method & method \cite{ma} & proposed algorithm \\
\noalign{\smallskip}\hline\noalign{\smallskip}
2 & 4 & 5 & 6 \\
4 & 24 & 25 & 36 \\
8 & 112 & 95 & 144 \\
16 & 480 & 325 & 480 \\
32 & 1984 & 1055 & 480 \\
64 & 8064 & 3325 & 1440 \\
128 & 32512 & 10295 & 4032 \\
256 & 130560 & 31525 & 10752 \\
512 & 523264 & 95855 & 69120 \\
1024 & 2095104 & 290125 & 168960 \\
\noalign{\smallskip}\hline
\end{tabular}
\end{table}

\section{Summary}
The article presents the novel algorithm for the DFRHT performing. The algorithm has a much lower computational complexity than the direct way of the DFRHT implementation. The computational procedure for DFRHT calculating is described in Kronecker product notation. The Kronecker product algebra is a very compact and simple mathematical formalism suitable for parallel realization. This notation enables us to represent adequately the space-time structures of an implemented computational process and directly maps these structures into the hardware realization space. For simplicity, we considered the synthesis of a fast algorithm for the DFRHT calculation for $N=2^3$. However it is clear that the proposed procedure was developed for the arbitrary case when the order of the matrix is a power of two. 




\end{document}